\documentclass[]{article} 
\usepackage{graphicx}
\usepackage{amssymb}
\usepackage{color}
\usepackage{slashed}
\usepackage{physics}
\usepackage[]{hyperref}
\usepackage[numbers]{natbib}
\usepackage{bibentry}
\usepackage{authblk}
\usepackage[offset=.38cm]{simpler-wick}

\usepackage[top=3cm, bottom=5cm, outer=3cm, inner=3cm]{geometry}

\hypersetup{
	colorlinks,
	linkcolor={red!50!black},
	citecolor={blue!50!black},
	urlcolor={blue!80!black}
}

\DeclareMathOperator*{\SumInt}{%
	\mathchoice%
	{\ooalign{$\displaystyle\sum$\cr\hidewidth$\displaystyle\int$\hidewidth\cr}}
	{\ooalign{\raisebox{.14\height}{\scalebox{.7}{$\textstyle\sum$}}\cr\hidewidth$\textstyle\int$\hidewidth\cr}}
	{\ooalign{\raisebox{.2\height}{\scalebox{.6}{$\scriptstyle\sum$}}\cr$\scriptstyle\int$\cr}}
	{\ooalign{\raisebox{.2\height}{\scalebox{.6}{$\scriptstyle\sum$}}\cr$\scriptstyle\int$\cr}}
}

\begin{document}
	
\title{The Role of Baryon Structure in Neutron Stars}
\author[1]{Theo F. Motta}
\author[1]{Anthony W. Thomas\footnote{Corresponding author}}
\affil[1]{CSSM and ARC Centre of Excellence for Dark Matter Particle Physics,
Department of Physics, University of Adelaide, SA 5005 Australia}

\maketitle

\newpage
\section{Introduction}
Arguably, the astrophysical result of most impact on the nuclear physics community of the last couple of decades was the precise determination that massive ($\gtrsim 1.9$M$_\odot$) neutron stars exist. The first measurement was made by Demorest {\it et al.}\cite{demorest2010two}, followed soon after by Antoniadis {\it et al.}\cite{Antoniadis2013} and others. The implications of the existence of such heavy stars for the nature of dense nuclear matter are numerous and these observations raised questions that still remain unanswered. 

The exact composition of the matter in the innermost parts of the core of such heavy stars is the subject of much debate and speculation. In spite of published work which had demonstrated that one could support such heavy stars even with hyperons in the core~\cite{RikovskaStone:2006ta}, Ref.~\cite{demorest2010two} suggested that their existence demonstrated the opposite. A plethora of models for the dynamics of dense matter that met the roughly $2$M$_\odot$ requirement rapidly appeared. These models ranged from exotic proposals of stars containing quark matter or strange matter to more conservative extensions of existing theories with additional, repulsive three-body forces, notably between hyperons and nucleons. 

On the observational side, the field has benefited enormously from the first gravitational wave (GW) measurement of a binary neutron star merger~\cite{ligoAbbott:2018exr}. Even though we have just a single example so far, the improved sensitivity that will follow from the current upgrade of the Laser Interferometer Gravitational-Wave Observatory (LIGO) suggests that we will see many more examples in the near future. The analysis of the signal from that one example has shown that we can use it not only to determine neutron star radii but in addition a completely new observable, the tidal deformability. Another major step forward has been the first data from the Neutron Star Interior Composition ExploreR (NICER)  mission~\cite{Riley:2019yda}, which promises to provide accurate measurements of neutron star radii.

There are now a number of models which provide a description of neutron stars and are not only consistent with the existence of heavy neutron stars but also lead to the presence of strange degrees of freedom in the core. The so-called hyperon puzzle, namely that the models that allow for the nucleons to decay to strange baryons do not agree with heavy mass measurements, even though hyperons are expected to be created given a large energy density, is naturally solved in these models. These attempts to describe dense matter with hyperon states require extra repulsion amongst such particles. We will review how such attempts are in fact consistent with the repulsive three-body forces that come from the change in the baryon structure in-medium. 

The structure of this review is the following. In the first part, we review the current astrophysical data. We then review those models which attempt to provide an account of the changes in baryon structure in dense media. We also review some attempts to model dense hyperonic matter via the introduction of extra phenomenological forces. Subsequently, we will discuss how these models succeed in providing a natural solution to the hyperon puzzle, alongside other possible solutions.
For other summaries of the nature of strange matter see, for example, the recent reviews \cite{Chatterjee:2015pua}, \cite{Gal:2016boi}, \cite{Tolos:2020aln} and \cite{Burgio:2021vgk}. For more detail on the equation of state of nuclear matter and its astrophysical implications, see \cite{Lattimer:2015nhk,Lattimer:2006xb,Ozel:2016oaf,menezes2021neutron,Orsaria:2019ftf,Baiotti:2019sew,Llanes-Estrada:2019wmz,Ma:2019ery,Stone:2021uju,Li:2021thg,Burns:2019byj,Dexheimer:2020zzs,Lattimer:2012nd,Weber:2004kj,Lattimer:2015nhk,Oertel:2016bki,Gandolfi:2015jma,Baym:2017whm,Chamel:2008ca,raduta2021eos}.

\section{Neutron Stars}
In this section, we provide a brief overview of the current data for masses and radii of neutron stars, where, for the latter, we restrict ourselves to the most recent NICER results and the GW constraints.

\subsection{Masses}
\begin{figure}[p]
	\centering
	\includegraphics[width=0.9\linewidth]{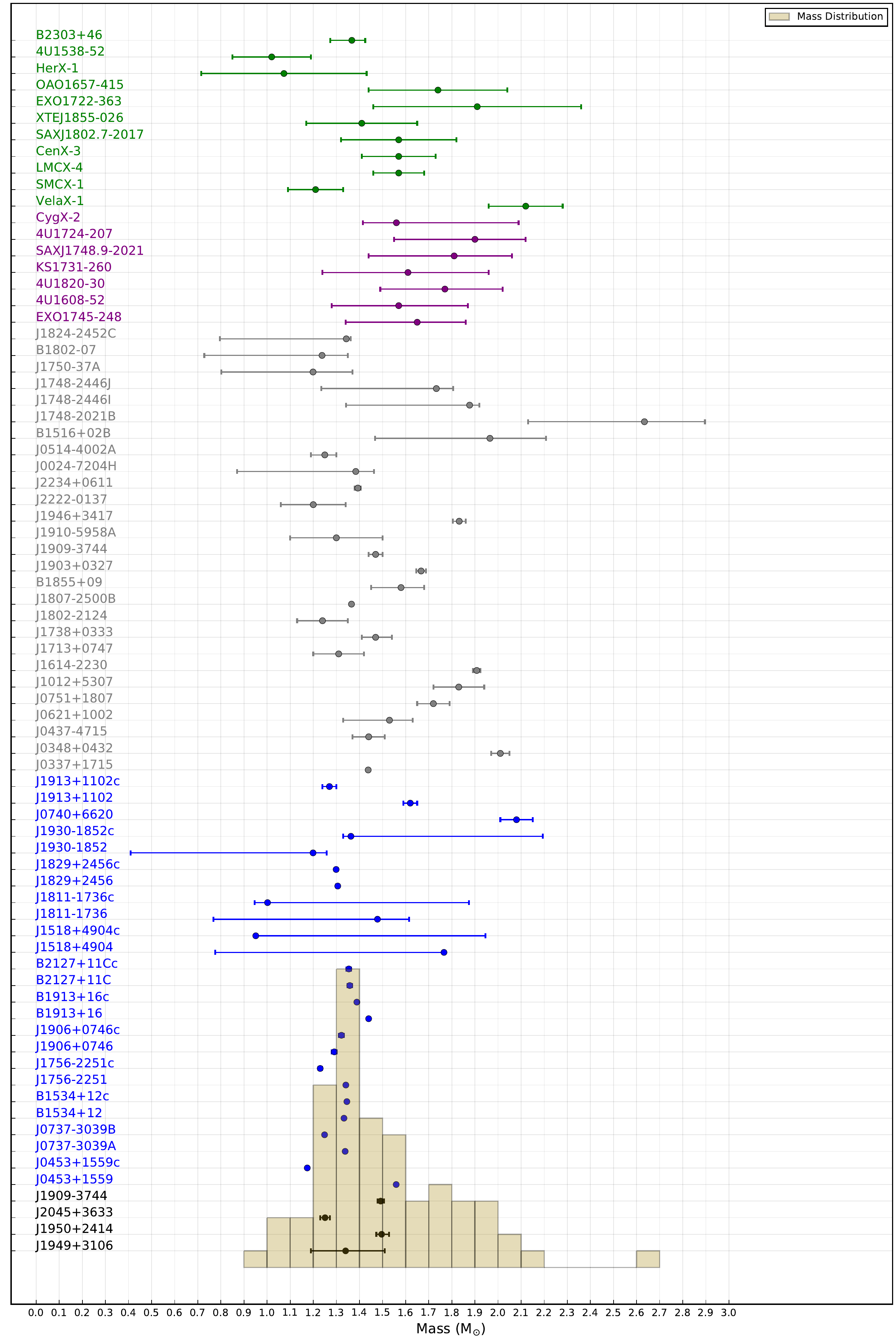}
	\caption{Neutron star masses gathered from \cite{Ozel:2016oaf} and updated with the measurements recently published in \cite{psr_ferdman2020asymmetric,psr_fonseca2021refined,psr_haniewicz2021precise,psr_liu2020revisit,psr_mckee2020precise,psr_ridolfi2019upgraded,psr_zhu2019mass,Nicer2Miller_2019}}
	\label{fig:mymasstable}
\end{figure}

The current state of neutron star mass measurements is depicted in Fig.~\ref{fig:mymasstable}. We see that the only measurements of heavy neutron stars with reasonably small errors are those of PSR J1614-2230 \cite{Arzoumanian} and PSR J0348+0432 \cite{Antoniadis2013}, along with PSR J0740+6620 \cite{psr_fonseca2021refined} which is the newest member of the heavy neutron star club. Nevertheless, the uncertainties on the aforementioned pulsars still make them all comparable with each other within two standard deviations. Therefore, a conservative lower estimate for the maximum mass of neutron stars is that of PSR J1614-2230, namely, 1.908$\pm 0.016$M$_\odot$.

\subsection{Radii}
Other than their masses, the radii of neutron stars are extremely sensitive to different models for the equations of state. However, the experimental constraints on radii are less numerous and until recently the systematic errors were quite large. The first new source of a constraint came on August 17th 2017, with the first gravitational wave measurement \cite{ligoAbbott:2018exr}  of the merger of two neutron stars. This yielded the result that the radii of the neutron stars involved in the merger both lie within $R=11.9^{+1.4}_{-1.4}$, at the 90\% confidence level.

More recently, the NICER mission, with its X-ray telescope on the space station, obtained results~\cite{Riley:2019yda,Nicer2Miller_2019} for PSR J0030+0451. The collaboration applied different analyses on the data and obtained values of (all at 68\% credibility) $12.71^{+1.14}_{-1.19}$km and mass $1.34^{+0.15}_{-0.16}$M$_\odot$ in Ref.~\cite{Riley:2019yda} and $13.02^{+1.24}_{-1.06}$km in Ref.~\cite{Nicer2Miller_2019} with the mass estimated at 1.44$^{+0.15}_{-0.14}$M$_\odot$. Another recent measurement by the NICER collaboration was that of pulsar PSR J0740+6620 which by virtue of being much heavier, is of considerable interest to nuclear physicists. They measured a radius of $12.39^{+1.3}_{-0.98}$km for a mass of $2.07\pm0.06$M$_\odot$ \cite{nicer2021}.

\section{Modelling Dense Matter}

Historically, studies of the properties of nuclear matter were primarily aimed at understanding finite nuclei. They were based primarily on using two-body and three-body potentials with standard techniques in quantum many-body theory, such as Brueckner-Hartree-Fock (BHF) \cite{goldstone}, Bethe-Brueckner-Goldstone (BBG) \cite{baldo2002bethe}. The two-body potentials were derived from phenomenological fits to nucleon-nucleon scattering~\cite{v18PhysRevC.51.38}. The quantum Monte Carlo variational approach has had considerable success with light nuclei~\cite{Wiringa:2000gb}, with parameters of the phenomenological three-body force tuned to nuclear data. 

In recent years there has been a great deal of activity using chiral effective field theory~\cite{Weinberg:1990rz,Weinberg:1992yk,Machleidt:2011zz,Drischler:2017wtt,bernard2008chiral,Greif:2020pju}, with its systematic expansion in powers of momentum. This approach builds in the constraints of chiral symmetry within a Lagrangian theory built upon pion and nucleon degrees of freedom, plus counterterms. With a similar number of parameters to those needed for phenomenological potentials (typically 25-30), it yields a good description of nucleon-nucleon scattering data. Within the same framework, one also has a systematic expansion of a three-nucleon force, with counterterms again tuned to nuclear data. Once again one finds excellent agreement with data for light nuclei~\cite{Freer:2017gip,Maris:2020qne}, with estimates of systematic errors. 

Building upon the work of Vautherin and Brink~\cite{Vautherin:1971aw}, the most popular theoretical treatment of finite nuclei has been the Hartree-Fock calculations using forces of the Skyrme type \cite{Skyrme:1956zz,Skyrme:1959zz,Chabanat:1997qh,Chabanat:1997un,Klupfel:2008af,Margueron:2012qx,Goriely:2015gzf,Goriely:2016sdz,Reinhard:2015ima,Dutra:2012mb,Tondeur:1984gwk}. In this approach one fits the parameters of the most general, local, velocity dependent energy functional to the properties of some set of finite nuclei~\cite{Stone:2006fn,Bogner:2013pxa}.

All of these non-relativistic methods are based on solid grounds and generally provide predictions in excellent agreement with their selected target nuclei. However, it is not clear that they can provide meaningful descriptions of neutron star matter much in excess of normal nuclear matter density. In particular, a non-relativistic approach cannot be considered reliable for densities beyond 1-2 times the saturation density of symmetric nuclear matter ($n_0$). Noting that the central density for the typical neutron star with mass 1.4 $M_\odot$ is of order $3 \, n_0$, one needs a different approach and this has led to the widespread application of Relativistic Mean-Field (RMF) models (the basis of which is reviewed in Ref.~\cite{walecka2004theoretical}). 

Of course, the RMF models are not without their difficulties. They often do not reproduce the properties of light nuclei as well as the non-relativistic models described earlier. The coupling constants in the relativistic Lagrangian density have to be fit to known nuclear matter parameters, however, there is often not enough data to constrain the potentially high number of parameters. This is especially so when one explicitly considers many-body forces, meson self-interactions and, of course, hyperons. For that reason, it is often difficult to gauge numerically the models' systematic biases and the number of models of the same family that provide often widely different predictions is high.

Lagrangian models are typically based upon a baryon-meson Lagrangian density, of which Eq.~(\ref{baryonlagrangian}) is a generic example. To this one can apply different techniques in many-body quantum field theory. Most commonly the model is treated in mean-field approximation (hence, relativistic mean-field). For instance, take the Lagrangian density
\begin{equation}
\label{baryonlagrangian}
\begin{aligned}
\mathcal{L}=&\bar\Psi^B\left(\slashed\partial - M_B + g_\sigma\sigma + g_\delta{\boldsymbol t\cdot\boldsymbol\delta} -g_\omega\slashed{\omega}-
g_\rho{\boldsymbol t\cdot{\boldsymbol{\slashed\rho}}}
\right)\Psi^B \\
&+\frac{1}{2}\left(\partial_{\mu} \sigma \partial^{\mu} \sigma-m_{\sigma}^{2} \sigma^{2}\right)
+\frac{1}{2}\left(\partial_{\mu} \boldsymbol\delta \cdot\partial^{\mu}  \boldsymbol\delta-m_{\delta}^{2} \boldsymbol\delta^{2}\right) \\
&-\frac{1}{4} \Omega_{\mu \nu} \Omega^{\mu \nu}+\frac{1}{2} m_{\omega}^{2} \omega_{\mu} \omega^{\mu} -\frac{1}{4} \boldsymbol{R}_{\mu \nu} \cdot \boldsymbol{R}^{\mu \nu}+\frac{1}{2} m_{\rho}^{2} \boldsymbol{\rho}_{\mu} \cdot \boldsymbol{\rho}^{\mu} .
\end{aligned}
\end{equation}
The static mean-field approximations would amount to
\begin{equation}
\begin{aligned}
\sigma &\rightarrow \expval{\sigma} = \bar\sigma \qquad
&\omega_\mu &\rightarrow \expval{\omega_\mu} = \expval{\omega_0} = \bar\omega \\
\boldsymbol{\rho_{\mu}} &\rightarrow \expval{\boldsymbol{\rho_{\mu}}} = \expval{{\rho_{0}}} = \bar\rho \qquad
&\boldsymbol{\delta} &\rightarrow \expval{\boldsymbol{\delta}} = \bar\delta \, .
\end{aligned}
\end{equation}
Under these approximations and with some trivial algebra, the problem of finding the Equation of State (EoS) of the system defined by the Lagrangian \ref{baryonlagrangian} can be solved \textit{exactly}, as it now amounts to a Fermi sea of dressed baryons
\begin{equation}
	\varepsilon_\text{Baryon} = \SumInt \frac{d^3 k}{(2\pi)^3} \sqrt{k^2 + M_B^{\star 2} } \, ,
\end{equation}
where the mass of the baryons is dressed by the mean scalar fields, and the meson mean-field component is
\begin{equation}
	\varepsilon_\text{Meson} = \frac{m_\sigma^2\bar\sigma^2}{2} 
	+\frac{m_\delta^2\bar\delta^2}{2}
	+\frac{m_\omega^2\bar\omega^2}{2}
	+\frac{m_\rho^2\bar\rho^2}{2} \, .
\end{equation}

The many different models in the RMF family are all based upon this foundation \cite{Walecka:1974qa,Chin:1977iz,Boguta:1977xi,Serot:1979dc,Typel:1999yq,Marques:2017zju,Sumiyoshi:1993wxt,Bender:2003jk,Ring:1996qi,Dutra:2014qga}. Modifications often include non-linear sigma potentials, $U(\sigma)$, and meson-meson couplings, such as $\sigma^2\delta$, $\sigma^2\delta^2$, $\rho^2\omega^2$ and the inclusion of explicitly density dependent parameters, amongst many others. Some more refined approaches also include the so-called exchange terms, or \emph{Fock} terms. This can be done as a perturbative correction to RMF or it can be included self-consistently (see for instance Refs.~\cite{Huber:1997mg,Miyatsu:2011bc}), in which case it is usually denoted Relativistic Hartree-Fock (RHF).

\section{Models Based on Quark-Meson Phenomenology}
As mentioned in the introduction, a particular sub-family of RMF models has proven effective in modelling finite nuclei, nuclear matter and neutron star core matter with a very small number of parameters. These models acknowledge that the baryon structure could play a role in nuclear dynamics. They are built upon the foundation of quark-meson phenomenology, which states that baryons in a nuclear medium can be modelled by a confined state of three quarks (interactions among which, such as one-gluon-exchange, can later be added perturbatively) that interact with quarks in other baryons by the exchange of meson fields.

In this section, we show a few examples of models in this family and discuss briefly their similarities and differences.

\subsection{The Quark-Meson Coupling Model (QMC)}
Physically, it is reasonable to assume that, the higher the density, and, as a consequence, the closer the baryons are to each other, the more likely it is that their internal degrees of freedom will play a role. Furthermore, considering the energy scales of the system, the scalar mean-field potential exerted on a nucleon \textit{at saturation density} is typically, depending on the model, hundreds of MeV, sometimes up to 0.5GeV (see for instance \cite{Brockmann:1990cn,walecka2004theoretical,Guichon:2018uew,Guichon:1995ue}). Such a strong scalar potential is comparable with the energy needed to excite a nucleon and indeed constitutes a large fraction of its mass, providing a strong indication that, even at saturation, the underlying structure of the baryon and its dynamics may play a role in many-body systems. However, attempting a full QCD solution of any nuclear-level phenomena is prohibitively difficult. It is reasonable to begin by modelling it phenomenologically.

The QMC model~\cite{Guichon:1995ue,Guichon:1987jp} attempts just that. Consider a single isolated baryon in a many-body system. Take it to be reasonably described by the MIT-Bag model, ergo, three light quarks trapped on a spherical cavity
\begin{equation}\label{MIT}\begin{aligned}
	&\mathcal{L}_0= \bar\psi_q (i \slashed\partial - m_q)\psi_q  -\mathcal{B}
	\quad \textrm{for} \ |\vec u |\leq R_B
	\\
	&(1+i\vec\gamma\cdot{\widehat u})\psi_q (\vec{u })=0 \quad \textrm{at} \ |\vec {u }| = R_B
\end{aligned}\end{equation}
where $R_B$ is the radius of the bag and $\mathcal{B}$ is the bag constant. The solution of this problem is elementary and expanding on the eigenstates of the free particle Hamiltonian
\begin{eqnarray}\begin{aligned}\label{MITbasis}
	&\left(
	-i\gamma^0\vec\gamma\cdot\nabla + \gamma^0 m_q
	\right)\phi(\vec{u })^\alpha = \frac{\Omega_\alpha}{R_B}\phi(\vec{u })^\alpha,\qquad
	&\int_{V_B}d^3u  \ \phi^i\phi^j=\delta^{ij},
\end{aligned}\end{eqnarray}
where $\alpha$ is a collection of quantum numbers, we obtain for a quark in its lowest energy state inside the bag,
\begin{eqnarray}\begin{aligned}
	\phi^{0m}(\vec{u })=\frac{\mathcal N}{\sqrt{4\pi}}\binom{j_0(xu /R_B)\chi_m}{-\beta_q\vec \sigma \cdot \hat{u } j_1(xu /R_B)\chi_m} \, ,
\end{aligned}\end{eqnarray}
where the boundary condition determines $x$ by $j_0(x)=\beta(x)j_1(x)$. The energy eigenvalue is just $\Omega_0=\sqrt{x^2+( m_qR_B)^2}$, the normalisation is $\mathcal{N}^{-2}=2R_B^3j_0^2(x)\left[ \Omega_0(\Omega_0-1)+ m_qR_B/2 \right]/x^2$, and finally, $\beta_q=\sqrt{\left(\Omega_0- m_qR_B\right)/\left(\Omega_0+ m_qR_B\right)}$.

However, as stated above, the model calculates the effects of the nuclear medium on this \textit{structure} (where structure is understood as the wave-function of the valence quarks inside the bag). The effective interaction between the \textit{light} quarks in different nucleons is mediated by the exchange of meson fields. 
These extra coupling terms are, for instance, with the scalar-isoscalar field $\sigma$, just $\mathcal L_I^{\sigma q} =g^q_\sigma  \sigma \bar\psi^{ }  \psi^{ }$, for the vector-isoscalar, $\mathcal L_I^{\omega q} =- g^q_\omega  \omega^\mu  \bar\psi^{ } \gamma_\mu \psi^{ }$, the vector-isovector and scalar-isovector, $\mathcal L_I^{\rho q} =- g^q_\rho \bar\psi^{ }  {\boldsymbol t\cdot\boldsymbol\rho^\mu} \gamma_\mu \psi^{ }$ and
$\mathcal L_I^{\delta q} =g^q_\delta \bar\psi^{ } {\boldsymbol t\cdot\boldsymbol\delta}  \psi^{ }$, where $\boldsymbol{t}$ denotes one half of the Pauli matrices for isospin.

In mean-field approximation, the scalar fields will modify the mass of the quarks, which, as can be seen by the expression for the wave function, will cause $\phi^{0m}(\vec{u})$ to depend on $\bar\sigma$ and $\bar\delta$ non-linearly. Since the strengths of the mean scalar fields depend on the scalar charge of the nucleons, one must solve the problem self-consistently at each density. On the other hand, the vector fields simply shift the energy of the quarks and thus, the energy of the bag in a linear way, with no change in the valence quark wave functions. This difference in the role of the scalar and vector mean-fields explains why it is not enough to just use the fact that the sum of the scalar and vector mean-fields is relatively small to dismiss any role for quark degrees of freedom in nuclei. 

Calculating the energy of a stationary bag, that is, the mass of a baryon, we obtain the result
\begin{eqnarray}\label{e0}
E_0=M_{N}^\star(\bar\sigma,\bar\delta) +\sum_q g_\omega^q\bar\omega
+\sum_q g_\rho^qI_q\bar\rho \, ,
\end{eqnarray}
with the nucleon effective mass
\begin{eqnarray}\label{massstar0}
M_{N}^\star(\bar\sigma,\bar\delta)=\frac{\sum_q\Omega_{0}^q(\bar\sigma,\bar\delta) N_{q} }{R_{B}}+\mathcal{B} V_{B} \, ,
\end{eqnarray}
where the index $q$ signifies the flavour of the quark. Before associating this directly with the mass of the baryon we may account for zero-point fluctuations and the centre of mass correction (namely $z_0$). Note that careful study of the centre of mass correction~\cite{Guichon:1995ue} showed that it is essentially independent of the mean scalar field, in contrast with early applications of the idea. One must also include one-gluon exchange, which provides a density-dependent colour hyperfine splitting~\cite{Guichon:2008zze} of otherwise degenerate baryons $\Delta E_M$. These modifications (see Refs.~\cite{RikovskaStone:2006ta,Guichon:2018uew}) yield a more detailed effective mass for the baryon
\begin{eqnarray}\label{massstar}
M_{N}^\star(\bar\sigma,\bar\delta)=\frac{\sum_q\Omega_{0}^q(\bar\sigma,\bar\delta) N_{q} - z_0 }{R_{B}}+\mathcal{B} V_{B} + \Delta E_{M}.
\end{eqnarray}

Note that not only the Dirac energy eigenvalues but also the OGE term depend on the mass of the quark, which depends on the scalar potentials. This in turn makes the effective mass of the baryons non-linearly dependent on the scalar mean-fields. This dependence, however, is not introduced \textit{a posteriori} with extra free parameters. Rather it is calculated within the model. Such non-linearity in the baryon effective mass is clearly a many-body effect and is equivalent to the introduction of many-body forces~\cite{Guichon:2004xg}; again with no new parameters. If we solve Eq.~(\ref{massstar}) for several values of the meson mean-fields we can deduce the functional dependence of the baryon mass on the scalar fields
\begin{eqnarray}\label{effmass2}
M_{N}^\star(\bar\sigma,\bar\delta)=M_B - g^B_\sigma(\bar\sigma,\bar\delta)\bar\sigma - g^B_\delta(\bar\sigma,\bar\delta)I_B\bar\delta \, ,
\end{eqnarray}
where $I_B$ is the isospin of the baryon. 

The way the effective mass depends on the mean-fields is clearly non-linear, however, it is accurately reproduced by a quadratic function of the isoscalar mean-field potential. That is, setting aside the much smaller effect of $\bar\delta$ for the moment, the effective mass is
\begin{equation}\label{effmass3}
    M^\star_N(\bar\sigma)= M_N - g_\sigma \bar\sigma + \frac{d}{2}(g_\sigma\bar\sigma)^2.
\end{equation}
The quantity $d$ is referred to as the scalar polarizability, by analogy with the electric polarizability. It describes the response of the internal structure of the nucleon which acts to oppose the applied scalar field. The dependence on the isovector-scalar potential is well approximated by a linear function, making the effective mass equal to~\cite{Motta:2019tjc}
\begin{equation}
    M^\star_N(\bar\sigma,\bar{\delta})= M_N - g_\sigma \bar\sigma + \frac{d}{2}(g_\sigma\bar\sigma)^2
    +g_\delta I_B \bar\delta - d_2\times(g_\sigma\bar\sigma)(g_\delta \bar\delta) \, ,
\end{equation}
to a very good approximation.

\subsubsection{Meson Structure}
It is true that the exchanged mesons have a quark structure just as much as the baryon. This is trivial for the $\omega$ and $\rho$ mesons, while the $\sigma$ is rather more complicated as it represents the exchange of two pions, including virtual $\Delta$ excitation, as well as the exchange of a quark-anti-quark composite. Therefore, some modifications of the QMC model take that into account as well~\cite{Saito:2005rv}. 

Let $\alpha$ be any meson field, then we can write its effective mass as
\begin{equation}
{m_\alpha^\star}=
m_\alpha +
a_1\sigma+
a_2\sigma^2+\cdots.
\end{equation}
Following Ref.~\cite{Saito:2005rv}, where the isovector-scalar meson was neglected, we discuss the meson mass dependence on the sigma mean-field only. The coefficients of this expansion can be free parameters, however, for the vector mesons one can make an expansion
\begin{equation}
	{m_\alpha^\star}=
	m_\alpha +
	\left(\frac{\partial m^\star_\alpha}{\partial\sigma}\right)_{\sigma=0}\sigma+
	\frac{1}{2!}\left(\frac{\partial^2 m^\star_\alpha}{\partial\sigma^2}\right)_{\sigma=0}\sigma^2+\cdots 
\end{equation}
and take the derivatives to be 2/3 of those for the baryon case -- given the quark content -- e.g. for the first derivative
\begin{equation}\label{mfmeson}
	\left(\frac{\partial m^\star_\alpha}{\partial\sigma}\right)_{\sigma=0}=\frac{2}{3}\left(\frac{\partial M^\star_N}{\partial\sigma}\right)_{\sigma=0} \, .
\end{equation}

However, some calculations using the Dyson-Schwinger equations have shown that the meson channel masses are more or less indifferent to the effects of chemical potential \cite{Gunkel:2019enr,Gunkel:2019xnh,Gunkel:2020wcl}. This could be explained by considering the quark-meson coupling model effects on the meson fields up to higher-order fluctuations. These fluctuation terms tend to increase the meson mass \cite{RikovskaStone:2006ta,Guichon:2006er} and hence tend to cancel the mean-field effects modelled in Eq.~(\ref{mfmeson}). For this reason, almost all applications of the QMC model have kept the meson masses independent of density.

\subsubsection{Other Effects}
Several other works have attempted to include additional effects associated with baryon structure.
One interesting modification is the inclusion of Fock terms at the quark level. Refs.~\cite{Nagai_2008,Miyatsu:2013exa,Miyatsu:2019xub} include such effects for pion exchange within the framework of the cloudy bag model, which we will discuss in more detail in Sec.~\ref{CQMC}. Including quark-level Fock terms results in a density dependence for the scalar polarizability (see Eq.~(\ref{effmass3})) and the dependence of the effective mass on the scalar potential is no longer well approximated by a parabola. This effect becomes even more non-linear if the quark-level Fock terms are included self-consistently, as in Ref.~\cite{Miyatsu:2010zzEffectOfGluon}. For the moment these calculations have not included the effect of short-range nucleon-nucleon correlations which significantly reduce short distance overlap between nucleons.

One can also include the Fock terms self-consistently at the baryon level, as done in Refs.~\cite{Miyatsu:2011bc,Miyatsu:2013hea,Whittenbury:2015ziz,Miyatsu:2020ujvTheRoleOfFock} and most recently  \cite{Stone:2021ngh}. According to Ref.~\cite{Miyatsu:2011bc}, the influence of the self-consistent Hartree-Fock approximation in the content and structure of a neutron star is sizable. The hyperon content of $\beta$-equilibrium matter in that work was restricted to the presence of $\Xi^-$ only and the maximum mass of the star went up substantially compared to the self-consistent Hartree calculation. The effects reported by Stone {\it et al.}~ \cite{Stone:2021ngh} were much less dramatic.

The fact that the baryon has a finite volume has also been considered. Aguirre {\it et al.}~\cite{Aguirre:2002xr} attempted to include excluded volume effects within the QMC model by imposing an equilibrium relation between the pressure of the bag and the baryonic pressure. They calculated the radius of the bag at each step in density and observed a lowering of the baryon size by 10\% in the star's centre compared to the vacuum case. 

\subsection{Structure of Finite Nuclei}\label{finitenuclei}
While the QMC model was formulated as a relativistic theory, for application to finite nuclei it is more convenient to use it to derive an equivalent, non-relativistic energy density functional (EDF)~\cite{Guichon:2006er}. This approach has become more sophisticated since the first systematic application by Stone et al.~\cite{Stone:2016qmi}. The latest version includes pion Fock terms~\cite{Stone:2017oqt}, a more accurate non-relativistic reduction of the $\sigma$ field in terms of density and the self-coupling of the $\sigma$ field of order $\sigma^3$~\cite{Martinez:2018xep}. The latter was found to be essential to reproduce the energies of the giant monopole resonances (GMR), with a nuclear incompressibility of order 240 MeV.
\begin{figure}[p]
	\centering
	\includegraphics[width=0.9\linewidth]{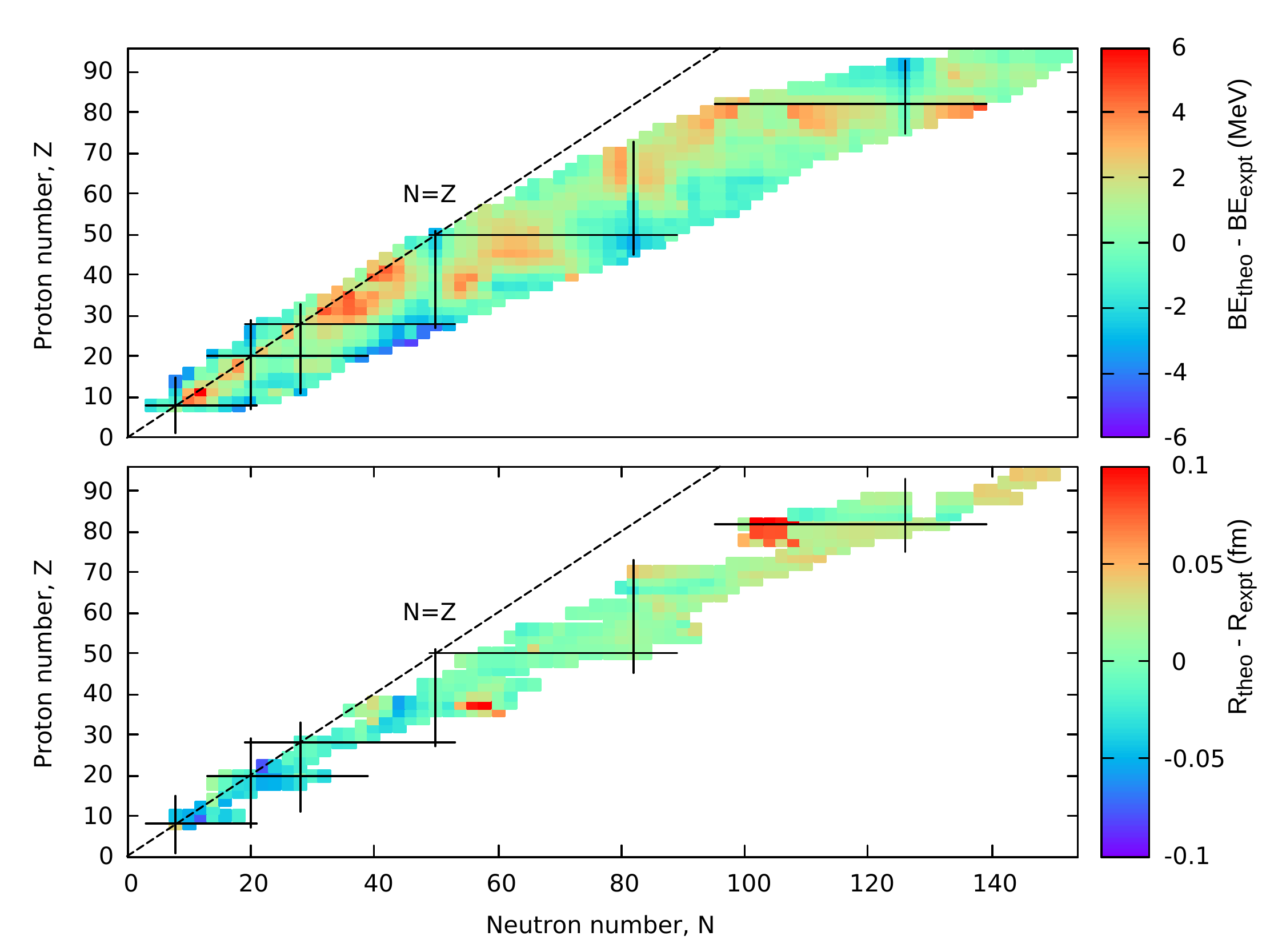}
	\caption{Deviations from experiment of the absolute binding energies and rms charge radii of all known even-even nuclei -- from Ref.~\cite{Martinez:2020ctv}.}
	\label{binding}
\end{figure}

Figure~\ref{binding} illustrates the level of agreement between the latest version of the model and the experimental binding energies and rms charge radii of all known even-even nuclei. The rms deviation of the binding energies is just 0.29\%, which is remarkable for a theory with just 5 parameters (the $\sigma \, , \omega$ and $\rho$ couplings to the light quarks, the mass of the $\sigma$ and the coefficient of the $\sigma^3$ interaction). For comparison UNDEF1~\cite{Kortelainen:2011ft,Bogner:2013pxa,Kortelainen:2010hv} reproduced the binding energies with a deviation of 0.55\%, while 
FRDM~\cite{Moller:2015fba}, with more than 25 parameters, managed 0.18\%. The rms charge radii were reproduced at the 0.5\% level, a small improvement on both UNDEF1 and DD-ME$\delta$~\cite{Typel:1999yq,Lalazissis:2005de}.
It is especially interesting that the binding energies of superheavy nuclei are reproduced at a level better than 0.1\%, which suggests the potential for application~\cite{Stone:2019syx} to the search for nuclei beyond Z = 118.

\subsection{Tests of changing hadron structure}
The argument that the large scalar mean-field in a nuclear medium may be expected to lead to changes in the structure of a bound proton is compelling~\cite{Thomas:2021kio}. Nevertheless it is critical to test this theoretical argument against experimental evidence. We briefly review the two most promising ways of testing this idea, which represents a fundamental change in paradigm for nuclear theory.

The first method to probe the structure of a bound nucleon involves the famous EMC effect~\cite{EuropeanMuon:1983wih,Geesaman:1995yd}. This revealed a significant reduction in valence quarks carrying a high momentum fraction in a nucleus compared with a free nucleon. The first calculation of this effect within the QMC model~\cite{Thomas:1989vt} revealed that it reproduced the major features of the effect. Later studies using the covariant NJL model, rather than the MIT bag model, for the basis of a QMC-like description of atomic nuclei, have demonstrated~\cite{Mineo:2003vc,Cloet:2005rt,Cloet:2009qs,Cloet:2012td} excellent agreement with modern data~\cite{Gomez:1993ri,Seely:2009gt} for the EMC effect across the periodic table.

In addition to describing existing data very well, the QMC approach predicts a number of new phenomena that can be tested in future experiments. This includes an isovector effect where the EMC effect is larger for $d$ than $u$ quarks in a nucleus with N$>$Z. This can be tested in parity violating deep inelastic scattering. It is also predicted that the spin dependent EMC effect, involving an unpaired valence proton, should be at least as large as the unpolarized EMC effect~\cite{Cloet:2005rt,Tronchin:2018mvu}. It also has the key advantage of discriminating~\cite{Thomas:2018kcx} between the explanations of the EMC effect based upon short range correlations~\cite{Weinstein:2010rt,CLAS:2019vsb} and those based upon the mean-field induced changes upon which we have focussed. This will be tested at Jefferson Lab on $^7$Li in the near future~\cite{JLabspin}.

\begin{figure}
		\centering
		\includegraphics[angle=0,width=0.6\textwidth]{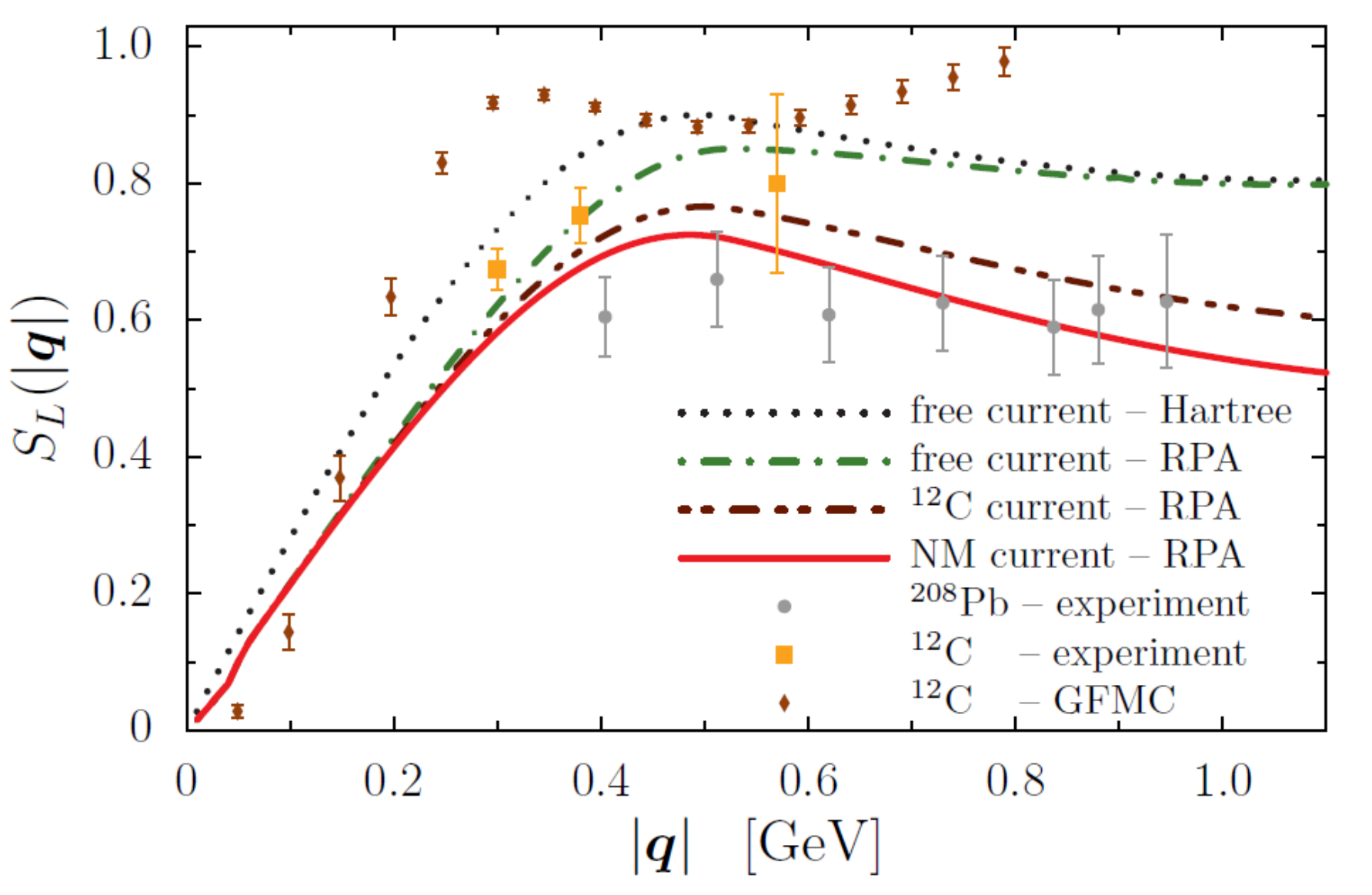}
		\caption{Coulomb sum rule determined at $\rho = 0, 0.1$ and 0.16 $fm^{-3}$, corresponding to a free nucleon current, a density typical of $^{12}$C; and nuclear matter saturation density. The data for $^{208}$Pb is from 
Refs.~\cite{Zghiche:1993xg,Morgenstern:2001jt} and for $^{12}$C from Ref.~\cite{Barreau:1983ht}, both without the relativistic correction factor of de Forest~\cite{DeForest:1984qe}. The Green's function Monte Carlo results are taken from Ref.~\cite{Lovato:2013cua}. The effects of relativity and the predicted modification of the proton electric form factor in-medium are both dramatic, with each tending to lower the Coulomb sum rule at large three-momentum transfer, $|\bf{q}|$, by as much as 20\%. RPA indicates that those calculations include the correlations arising within the random phase approximation. (The figure is taken from Ref.~\cite{Cloet:2015tha}.) }
\label{fig.Coulomb}
	\end{figure}

An additional key test of the idea stems from remarkable early work on the Coulomb sum-rule by Morgenstern, Meziani and collaborators~\cite{Barreau:1983ht,Morgenstern:2001jt}. This work strongly indicated that the electric form factor of a bound proton was significantly different from that of a free proton, a result expected within the QMC model~\cite{Lu:1997mu,Cloet:2015tha}. Precise new results from Jefferson Lab, which are expected to resolve this issue experimentally, are under analysis at the present time. The current theoretical and experimental situation is summarized in Fig.~\ref{fig.Coulomb}, which is taken from Ref.~\cite{Cloet:2015tha}.

\subsection{The Bogoliubov-QMC Model (B-QMC)}
An interesting recent proposal~\cite{Bohr:2015fgc,Panda:2018tax} that also takes into account in-medium changes in baryon structure is based on a Bogoliubov Hamiltonian with quark-meson interactions in mean-field approximation. That is, the quarks are described by eigenstates of %
\begin{equation}
	h_{D}=-i \boldsymbol{\alpha} \cdot \nabla+\beta\left(\kappa|\mathbf{r}|+m-g_{\sigma}^{q} \sigma\right) \, .
\end{equation}
The most obvious fundamental difference between this model and the basic QMC model described in the previous section is the fact that here~\cite{Bohr:2015fgc}, what confines the quarks is a linear potential rather than an infinite spherical well. The free parameter $\kappa$, which is essentially the string tension, was fixed so as to reproduce the mass of the free nucleon and, as previously with the QMC model, an energy density was derived from an RMF construction where the mass of the nucleon was the effective mass calculated by the model. We shall return to this model in section~\ref{hypcr}, however, it is worth pointing out that it also predicts heavy stars with strange baryonic matter in the core. The predictions for hyperonic stars go up to $2$M$_\odot$ and the species fraction obtained in beta-equilibrium is similar to the QMC result. The indication here seems to be that the results are somewhat independent of the confinement model (a hypothesis that is also supported by a third model we will discuss in the next section).

Much has happened, however, since the publication of these original papers. The most recent discussion of the model~\cite{Rabhi:2020ggz} includes an $s\bar s$ component in the vector-isoscalar channel, constrained by $\Lambda$ hypernuclear data. We shall return to this point in section~\ref{hypcr}.

\subsection{The Quark-Mean-Field Model (QMF)}
A further example of a model that is based on quark-meson interaction but nevertheless attempts to provide a different description of the confining potential is the Quark-Mean-Field model~\cite{Li:2020dst,Mishra:2015hyy,Mishra:2016qhw}. In this work the confining potential is taken to be a harmonic oscillator, which is an equal mixture of Lorentz scalar and vector terms. This has the property of being analytically solvable in the Dirac equation
\begin{equation}
	U(r)=\frac{1}{2}\left(1+\gamma^{0}\right)\left(a r^{2}+V_{0}\right) \, .
\end{equation}
%
As with the previous two models mentioned, they write an RMF model from the solution to the Dirac equation, designed to reproduce the appropriate effective mass that their microscopic model of the baryon yields. They also verify hyperonic matter able to reach the 2M$_\odot$ limit.

\subsection{Chiral and Cloudy QMC}\label{CQMC}
Within Ref.~\cite{Nagai_2008} and later in Refs~\cite{Miyatsu:2013exa,Miyatsu:2019xub} the Cloudy Bag Model (CBM) \cite{Miller:1979kg,Thomas:1981vc,Thomas:1982kv,Miller:1984em} was used within the framework of the quark meson phenomenology. Taking a typical CBM Lagrangian density, such as
\begin{equation}
	\mathcal{L}_{\text {CBM }}=\left[\bar{\psi}\left\{i \gamma_{\mu} \partial^{\mu}-m+i \frac{m}{f_{\pi}} \gamma_{5} \vec{\tau} \cdot \vec{\phi}+\frac{1}{2 f_{\pi}} \gamma_{\mu} \gamma_{5} \vec{\tau} \cdot\left(\partial^{\mu} \vec{\phi}\right)\right\} \psi-B\right] \theta_{V}-\frac{1}{2} \bar{\psi} \psi \delta_{S}+\frac{1}{2}\left(\partial_{\mu} \vec{\phi}\right)^{2}-\frac{1}{2} m_{\pi}^{2} \vec{\phi}^{2}
	\, ,
\end{equation}
where $B$ is the bag constant, $\theta_V$ the bag volume step function and $\delta_S$ the Dirac delta function at the bag surface, and finally $\phi$ is the pseudo-scalar pion field. In particular, Ref.~\cite{Nagai_2008} also includes gluon exchange between the quarks. Here, as opposed to the classical QMC model, the energy of the stationary bag -- the mass of the baryon -- is calculated in Hartree-Fock approximation \emph{at the quark level} (whereas usually the Fock terms tend to be neglected) and finite size effects for the pion are also included via a form factor for the pion-quark interaction. This results in a more complex density dependence of the mass of the baryon and, particularly, a density-dependent scalar polarizability (the $d$ factor in Eq.~(\ref{effmass3})). 

After including the pion cloud and gluon exchange between the quarks, one then proceeds to calculate the equation of state at the baryonic level. Taking this effective mass, $M^\star$, Ref.~\cite{Nagai_2008} includes $\sigma$ and $\omega$ exchange between the baryons in mean-field approximation. That is, Fock terms are omitted at the baryon-meson level, and the effects of NN correlations are not included, for example by dropping the contact terms in the pion-nucleon Fock terms, as done in Refs.~\cite{Stone:2006fn,Guichon:2018uew} and others. Miyatsu {\it et al.}~\cite{Miyatsu:2019xub} do, however, add the isovector-vector $\rho$ exchange as well as the strange mesonic channels, $\sigma^\star$ and $\phi$, which were missing in Ref.~\cite{Nagai_2008}.

As shown in \cite{Miyatsu:2019xub}, the gluon interaction tends to stiffen the EoS and the full CBM based quark-meson coupling model does tend to give higher maximum masses for neutron stars compared with the standard QMC model by almost $0.2$M$_\odot$. However, it also tends to increase the incompressibility of nuclear matter by a significant amount, putting it at $K_0\approx 310$MeV, which is higher than most giant monopole resonance data suggests it should be~\cite{Martinez:2018xep,Youngblood:1999zza,Garg:2018uam}. 

\subsection{The NJL Based QMC Model}
When it comes to the phenomenology of dense matter, no model has been more prolific than the NJL model (an excellent review of which can be found in Ref.~\cite{Buballa:2003qv}). Most of its applications have been to model deconfined quark matter (see Ref.~\cite{Buballa:2003qv}). However, there have been a few works on baryon structure ranging from hadronisation \cite{PhysRevC.89.065204} to baryon dissociation in hot \cite{PhysRevC.91.065206} and dense \cite{WANG2011347} media.
The NJL model has been successful describing hadron properties in vacuum. Including an infra-red cutoff within the proper-time regularisation scheme in the NJL model simulates confinement, in that the quarks cannot be on-mass-shell, and it can also be used to derive~\cite{Bentz:2001vc} a quark-meson phenomenological model. Refs.~\cite{Lawley:2004bm,Lawley:2006ps} were devoted to such a calculation. Starting from an $SU(2)$-flavour symmetric quark Lagrangian 
\begin{equation}
\mathcal{L}=\bar{\psi}(i \slashed \partial-m) \psi+\sum_{\alpha} G_{\alpha}\left(\bar{\psi} \Gamma_{\alpha} \psi\right)^{2} \, ,
\end{equation}
in the same spirit as the QMC model, one calculates the effective mass of the baryon in medium, 
\begin{equation}
M=m - 2G_\pi \bra{\rho} \bar\psi\psi \ket{\rho } \, , 
\end{equation}
and implements it within a baryon-level calculation, in the case of \cite{Lawley:2004bm} using the Lagrangian density
\begin{equation}
\mathcal{L}= \bar{\psi}\left(i\slashed\partial-M-2 G_{\omega} \gamma^{\mu} \omega_{\mu}-G_{\rho} \gamma^{\mu} \tau \cdot \rho_{\mu}\right) \psi-\frac{(M-m)^{2}}{4 G_{\pi}}+G_{\omega} \omega^{\mu} \omega_{\mu}+G_{\rho} \rho^{\mu} \rho_{\mu}+\mathcal{L}_{I} \, .
\end{equation}

\begin{figure}
	\centering
	\includegraphics[width=0.7\linewidth]{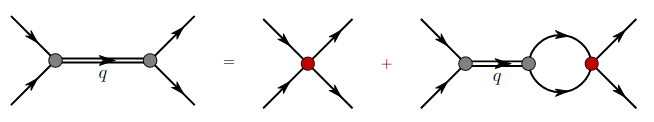}
	\caption{Bethe-Salpeter equation used in Ref.~\cite{Whittenbury:2016sma}}
	\label{fig:bethes}
	\centering
	\includegraphics[width=0.6\linewidth]{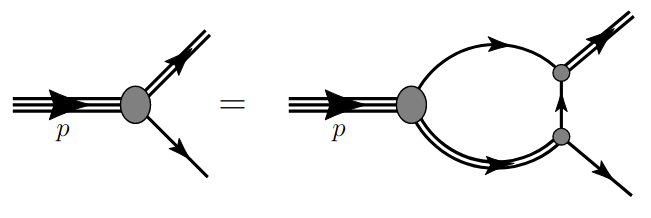}
	\caption{Fadeev equation describing quark-diquark correlations from the same reference}
	\label{fig:fadeev}
\end{figure}

A further attempt to substitute the bag model description of the baryon with an alternative model was carried out in Ref.~\cite{Whittenbury:2016sma} where, within the NJL model, based on the fundamental work of Ref.~\cite{Bentz:2001vc}, one self-consistently solves a Bethe-Salpeter equation for the diquarks (Fig.~\ref{fig:bethes}) and a Faddeev equation for the baryons described as quark-diquark correlations (Fig.~\ref{fig:fadeev})

In Ref.~\cite{Whittenbury:2016sma} it was once again found that the mass of the baryons depends non-linearly on the sigma mean-fields and, here again, it was found that this change affects the equation of state and the composition of dense matter. Importantly, as in the previous models which showed that the $\Sigma$ baryon is energetically unfavourable and thus does not contribute, here too the $\Sigma$ was found to be unbound.
Nevertheless, it was found that the equation of state within this model is softer in comparison with the other quark-meson based models mentioned above, including other NJL efforts such as Refs.~\cite{Lawley:2004bm,Whittenbury:2013wma}. In Ref.~\cite{Whittenbury:2015ziz} the NJL approach of \cite{Whittenbury:2016sma} was extended to include a deconfined quark phase under the same model.

\section{RMF Models with Many-Body Forces}
In the previous sections we have seen a number of models which share some common features. They all attempt to provide a dense matter description that takes into account the fact that the baryon is not a fundamental object. However, it is noticeable that they are all based on very similar grounds, the phenomenological quark-meson models. Let us approach other RMF-type models that also succeed in providing a maximum mass of order $2$M$_\odot$ result \textit{with} hyperon content but which are not based on the quark-meson phenomenology.

Some models, such as Refs.~\cite{Gomes:2015fsa,Gomes_2015,Gomes:2018jmu,Vasconcellos:2014qua} and \cite{razeira2014strangeness}, amongst others, attempt to provide an RMF description with phenomenological many-body forces. Recognising that such forces exist and must play an essential role, Ref.~\cite{Gomes_2015} provides a claim for the effective coupling of the scalar mesons to the baryons, namely
\begin{equation}
\begin{aligned}
	g_{\sigma b}^{*} \equiv\left(1+\frac{g_{\sigma b} \sigma+g_{\delta b} I_{3 b} \delta_{3}}{\zeta m_{b}}\right)^{-\zeta} g_{\sigma b},\\ g_{\delta b}^{*} \equiv\left(1+\frac{g_{\sigma b} \sigma+g_{\delta b} I_{3 b} \delta_{3}}{\zeta m_{b}}\right)^{-\zeta} g_{\delta b},
\end{aligned}
\end{equation}
where $\zeta$ is a model parameter. This is not at all based on any claim about the internal hadron structure, however, by virtue of its non-linearity, this inevitably yields many-body forces and is, in fact, comparable to the QMC model. What it does that the QMC model does not is to allow for different non-linear behaviour of the coupling, subject to this free parameter, $\zeta$. The QMC model, in comparison, provides a clear origin for these many-body forces, with no additional parameters, based upon the medium induced baryon structure alterations.

\begin{figure}[t!]
	\centering
	\includegraphics[width=0.9\linewidth]{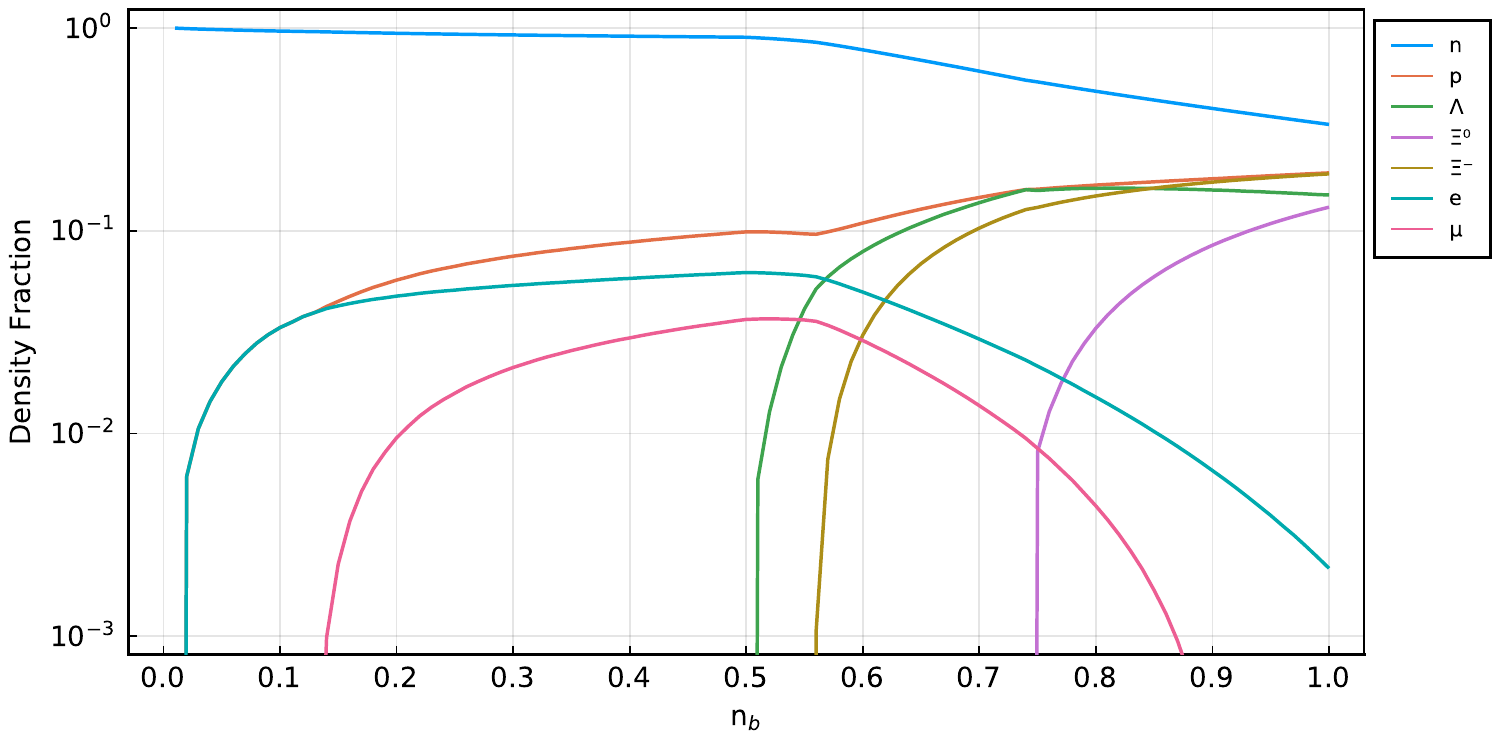}
	\caption{Species fraction in the QMC model}
	\label{fig:speciesfrac}
\end{figure}

That this approach is, to some extent, phenomenologically similar to the QMC model and its variations is best and most easily shown in the species fraction. Ref.~\cite{Gomes:2015fsa}, for instance, finds that the $\Sigma$s are suppressed and do not appear in dense nuclear matter (since we know from data on finite nuclei that the $\Sigma$N interaction must be repulsive~\cite{Dover:1989sv,Tsushima:2009zh}, that is correct) and it also has the $\Lambda$ appear before the $\Xi$. Fig.~\ref{fig:speciesfrac} shows the same in the QMC model.

The benefit of the purely phenomenological approach is that it allows for a lot of flexibility on the functional dependence of the effective couplings with the mean-field. However, the benefit of the quark-meson approach is that it provides a simple explanation for the nature of these non-linear couplings without new parameters. Indeed it was the only theory to anticipate the presence of hyperons in heavy neutron stars {\em before}  they were discovered~\cite{RikovskaStone:2006ta}. Nevertheless, both approaches seem to agree on a number of key observables.

\section{The Strange Puzzle}\label{hypcr}
One common trend amongst the quark-meson type models is that they all naturally include many-body forces that increase the pressure of infinite nuclear matter with hyperon degrees of freedom. When it comes to the so called hyperon puzzle, moreover, they all -- or most of them -- predict heavy neutron stars with maximum masses of order 
2 M$_\odot$, often with fewer parameters as well. Furthermore, when used to derive an equivalent non-relativistic energy density functional, the QMC model also provides excellent agreement with data on finite nuclei with a comparatively small number of parameters \cite{Stone:2016qmi,Martinez:2020ctv}.   

The underlying description of nuclear matter in the quark-meson approach is fully relativistic and completely consistent with data both from astrophysical and terrestrial experiments. Nevertheless, there \textit{are} puzzles about the composition of the matter inside neutron star cores. While the terminology of a ``hyperon puzzle'' is clearly outdated, given that several models that include the full octet are consistent with the data (quark-meson models or otherwise), whether the strange matter is confined or deconfined \textit{is} truly a puzzle. This section is devoted to the investigation of such a puzzle.

\subsection{Modified Gravity as a Solution}
From the discovery of the mass of PSR-J1614 by \cite{demorest2010two}, the notion that models of neutron star matter that include hyperons in the core were in tension with heavy masses was widespread (in spite of the published work in Ref.~\cite{RikovskaStone:2006ta}). Amongst the first few attempts to solve this issue after the work of Demorest {\em et al.},  Refs.~\cite{Cooney:2009rr,Yunes:2009ch,Ali-Haimoud:2011zme,Yin:2011aa,Deliduman:2011nw,Orellana:2013gn,Zhang:2013cca,Yagi:2013mbt,Cheoun:2013tsa,Yunes:2013dva,Alavirad:2013paa,Astashenok:2013vza,Astashenok:2014pua,Sham:2013cya,Sotani:2014goa,Glampedakis:2015sua,Capozziello:2015yza,Sakstein:2016oel,Burrage:2017qrf} put forward the idea that this puzzle is not a crisis of the nuclear equation of state physics, but rather of gravitation. Alternative theories to Einstein's General Relativity (GR) or, modified gravity, are alternative hypothesis to dark matter in solving open cosmological questions. These works continue to the present day \cite{Lope-Oter:2021vxl,Prasetyo:2021kfx,Azri:2021pxu}, and, although alternative ways to deal with the hyperon problem from the nuclear perspective have been found, it remains an important hypothesis. A good review of such theories can be found in Ref.~\cite{Hess:2020ssc}.

\subsection{Many-Body Forces}
As discussed in a previous section, Refs~\cite{Gomes:2015fsa,Gomes_2015,Gomes:2018jmu,Vasconcellos:2014qua} all attempt to include many-body forces directly within an RMF description of nuclear matter. Importantly, though, the strangeness puzzles are also of interest to $\chi$EFT practitioners. The recent  Refs.~\cite{Haidenbauer:2016vfq,Logoteta:2019utx,	Gerstung:2020ktv,Logoteta:2017asv} all pay attention to these $\Delta$ and hyperon many body forces from a $\chi$EFT perspective.
For instance, Ref.~\cite{Logoteta:2019utx} employs a three body hyperon interaction derived by the
Jülich-Bonn-Munich collaboration \cite{jbmPhysRevC.93.014001}. They find that the NN$\Lambda$ repulsion pushes the Lambda appearance threshold to higher densities as many body forces usually do \cite{Motta:2019tjc,Motta:2019ywl,Stone:2006fn,Guichon:2018uew,Gomes:2015fsa,Gomes_2015,Gomes:2018jmu,Vasconcellos:2014qua}. The influence of this many-body repulsion on the equation of state and the neutron star mass is huge, taking the maximum mass from 1.3 to above 1.9M$_\odot$.

From the perspective of other approaches to microscopic potential models, Ref.~\cite{Rijken:2016uon} employs a model for hyperon interactions based on meson, pomeron and odderon exchanges, with which they find low masses for neutron stars -- a common issue amongst models that include hyperons (the aforementioned hyperon puzzle).

Temperature effects and heat transport coefficients are also of great interest. 
In the recent works of Refs.~\cite{Lu:2019mza,Tolos:2019wea,Logoteta:2020yxf}, temperature effects were included within the BHF formalism. Particularly Ref.~\cite{Tolos:2019wea} shows both models with hyperons and without hyperons agreeing with cooling observations. In fact, the constraints on neutron star thermal evolution -- or more specifically the cooling speed -- were also recently taken into account in Refs.~\cite{Wei:2020saq} and \cite{Wei:2020xdk}, while in \cite{Shternin:2017rlt,Shternin:2020igy} the effects of three-body forces were taken into account in BHF theory at finite temperature. Finally, we note that a complete treatment of Fock terms within the QMC model at finite temperature has recently been reported by Stone and collaborators~\cite{Stone:2021ngh}.

\subsection{Strange Mesons}
Another important piece of this puzzle is, what is the role of strange mesons? Some early works which were particularly impactful \cite{Suzuki:1976tj,Haensel:1982zz,Ellis:1995kz,Tsushima:1997df,Pal:1999sq} deal with this question, as well as several subsequent publications \cite{Kolomeitsev:2017gli,Chamel:2017wwp,Miyatsu:2015kwa,Maslov:2015msa,Menezes:2005ic,Ryu:2006xx,Ryu:2007qz,Ryu:2011ny}. These works attempt to address the aforementioned issue. Particularly, Ref.~\cite{Maslov:2015msa} shows that their RMF model, in the presence of a strange $\phi$ mean-field, recovers the 2M$_\odot$ threshold and is compatible with available data. 

One important issue in neutron star physics is that of meson condensation, including kaon condensation. This has been studied, for example, in Refs.~\cite{Tsushima:1997df,Menezes:2005ic,Ryu:2006xx,Ryu:2007qz,Ryu:2011ny,Maruyama:2005yp}. The origin of the interest in kaon condensation is the early idea of \cite{migdal1973pi} pion condensation in nuclei. This was based on the proposal that, at a certain density, nuclear matter would suffer a phase transition into a phase with a non-zero pion condensate. The threshold for such a transition would be the point at which pions can be created at no energy cost. Several models for the pion self-energy in nuclear matter attempted to pinpoint such a density threshold (see Ref.~\cite{Ericson:1988gk} and explained the semi-crystalline structure that nuclear matter would have to assume for such a condensate to be possible \cite{takatsuka1976note,tamagaki1976pi0}. 

It has been shown, however, that short range repulsive interactions such as omega exchange is enough to prohibit pion condensation and the idea has more or less been dismissed. However, both pion and kaon condensation (via a nearly identical mechanism) were proposed to exist at densities far higher than nuclear matter, perhaps being stable in the core of neutron stars. Within the framework of quark-meson phenomenological models this hypothesis was studied in Refs.~\cite{Tsushima:1997df,Menezes:2005ic,Ryu:2006xx,Ryu:2007qz,Ryu:2011ny}. The quark-meson framework was implemented in a similar way as it is for the baryon states, where the effective mass of the hadron is calculated via the bag model (or whatever other confining model) for non-interacting quarks that couple to $\sigma$, $\omega$ and $\rho$. The mass of the kaon is derived naturally as \cite{Tsushima:1997df}
\begin{equation}
	m_{K}^{\star}=\frac{\Omega^{\star}+\Omega_{s}-z_{K}}{R_{K}^{\star}}+\frac{4}{3} \pi R_{K}^{\star 3} B \, ,
\end{equation}
where its radius is fixed by $\partial_{R_K} m^\star_K =0$, and one uses the effective Lagrangian 
\begin{equation}
	\mathcal{L}=\left[\left(\partial_{\mu}+i g_{\omega}^{K} \omega_{\mu}+i g_{\rho}^{K} \frac{\tau_{3}}{2} \rho_{\mu}\right) K\right]^{\dagger}\left[\left(\partial^{\mu}+i g_{\omega}^{K} \omega^{\mu}+i g_{\rho}^{K} \frac{\tau_{3}}{2} \rho^{\mu}\right) K\right]-m_{K}^{\star 2} \bar{K} K+\mathcal{L}_{\text {matter }} \, .
\end{equation}
Once the kaon density is incorporated into the equation of state, stellar structure equation solutions show that, as expected, the tendency is for the kaon condensation to lower the maximum mass. This would ``solve'' the hyperon puzzle as it accounts for strange degrees of freedom, which are expected at high density, but does not suffer from such a drastic softening of the EoS.

Another solution to the hyperon puzzle that involves strange mesons is to include the $\phi$ and $\sigma^\star$ mesons (e.g.,  Refs.~\cite{Pal:1999sq,Maslov:2015msa,Miyatsu:2015kwa}, amongst others). The extra repulsion of the $\phi$ meson increases the stiffness of the EoS. Together with a self-consistent implementation of the $\sigma^\star$ which, as is typical of quark-meson coupling models, produces attraction that grows less quickly than the density because of the scalar polarizability (see Eq.~(\ref{effmass3})), also comfortably solves the so called hyperon problem.

However, it is important to be careful with respect to the explicit introduction of extra strange degrees of freedom. Constraining the coupling constants at saturation density is difficult enough with non-strange mesons. Perhaps most importantly, typical values of the $\sigma$ and $\omega$ coupling constants when applied to hyperons reproduce the gross properties of $\Lambda$ hypernuclei. 

\subsection{Exotic Ideas}
There are more hypotheses, however, that do not neatly fit this dichotomy between hyperons or quarks in the core. One prominent hypotheses is the so called Quarkyonic matter \cite{mclerran2019quarkyonic}. According to this hypothesis, dense QCD matter would be comprised of deconfined quark matter. However, quantum fluctuations at and above the Fermi surface would manifest only as colourless states, effectively creating a layer at the top of the Fermi sea which would be populated by baryons. Both quarks and baryons would populate the core and would be discriminated only by momentum. For detailed discussions on quarkyonic matter and its possible role in neutron star physics see Refs~\cite{mclerran2019quarkyonic,philipsen2018baryonic,zhao2020quarkyonic,sen2021finite,mclerran2009quarkyonic,cao2020field}. 

One key feature of the quarkyonic model that makes it interesting is that it exhibits a characteristic peak in the speed of sound at mid-range densities, after which the speed of sound goes down again. This early stiffness helps the star to sustain higher masses. However, it does seem to be somewhat in contrast with the claims of Ref.~\cite{Annala:2019puf} in which lower maximum speed of sound EoSs (specially those which do not grossly violate the conformal limit of $c_s^2=0.3$) are correlated with higher neutron star masses.

Other hypotheses also contemplate the possibility that both the residual strong force of baryons interacting with baryons by the exchange of meson fields, and the fundamental strong force of quarks and gluons could play a role simultaneously. One example is the so called Multi-Pomeron exchange Potential (MPP)
\cite{Yamamoto:2017sdb,yamamoto2013multi,yamamoto2014hyperon,yamamoto2016hyperon,yamamoto2017neutron,lim2018effective,Logoteta:2019utx,blaschke2021studying} according to which there could be, at high densities, a pomeron based many-body repulsion that, by virtue of being universal amongst all baryons, would not affect the baryon content of the star but could generate a large amount of repulsion. 

Another example of a model that takes into consideration quarks and baryon degrees of freedom together is the quark percolation model
\cite{GUTTNER1986555,fukushima2020hard,kiguchi1981neutrino,Masuda:2012ed}.
This is based on the premise that at sufficiently high densities the distinction between the baryonic picture and quarkyonic picture becomes blurred. If the density is high enough the baryon wave functions would start to overlap. The quarks in the meson clouds of the overlapping baryons would behave as sea quarks and their wave functions would become constant in space. 

All of these hypotheses -- quarkyonic matter, the MPP model, and the percolation model -- are based on very different ideas. Nevertheless, they are similar inasmuch they discuss the physics of regimes where the relevant degrees of freedom are neither exclusively quarks and gluons nor exclusively baryons and mesons.

\subsection{Strange Quark Matter or Hyperonic Core?}
The notion of a ``hyperon'' puzzle as it was first conceived is certainly outdated. Several models are capable of obtaining heavy stars with hyperon cores, as discussed extensively in this review. However, the puzzle that \textit{does} remain is whether the strange matter in the core of neutron stars is confined or deconfined?
The hypothesis of quark matter in the core of neutron stars is old \cite{Collins:1974ky,Baym:1976yu} and has gained much traction recently, especially with the work of Annala {\em et al.}~\cite{Annala:2019puf}. These authors devised a model-independent parametrization of the EoS via a linear piecewise construction of the speed of sound (squared) of dense matter as a function of the baryonic chemical potential:
\begin{equation}
	c_{\mathrm{s}}^{2}(\mu)=\frac{\left(\mu_{i+1}-\mu\right) c_{\mathrm{s}, i}^{2}+\left(\mu-\mu_{i}\right) c_{\mathrm{s}, i+1}^{2}}{\mu_{i+1}-\mu_{i}}
\end{equation}
for a given set
\begin{equation}
	\left\{\left(\mu_{i}, c_{\mathrm{s}, i}^{2}\right)\right\}_{i=1}^{N_{p}} \, .
\end{equation}
From the speed of sound one can obtain the full equation of state via standard thermodynamic relations.

Based on the parametrization of the speed of sound they generated a large set of EoSs via random Monte Carlo, constraining them at large density by perturbative QCD and low densities by $\chi$EFT, and applied certain selection rules on these functions. Causal EoS that yield stars with a mass at least $2$ M$_\odot$ were shown to have a common behaviour, with a speed of sound that approached the conformal value $c_s^2=1/3$ and polytropic index less than $\lesssim1.75$. One possible interpretation is that this is evidence of quark matter \cite{Annala:2019puf}, given that very high density quark matter in the perturbative regime shows these characteristics. 

\begin{center}
	\begin{figure}
		\includegraphics[scale=0.5]{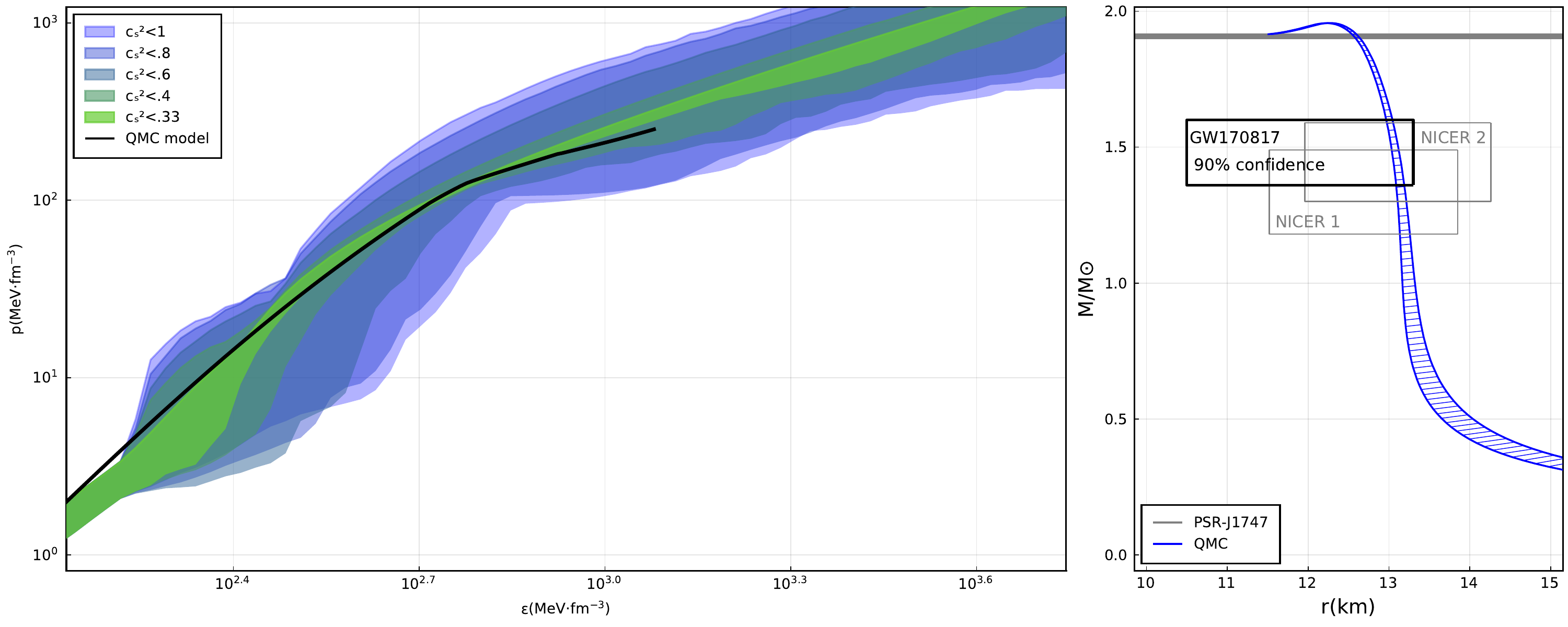}
		\caption{Allowed EoSs band from Ref.~\cite{Annala:2019puf} overlapped with the QMC model EoS from Ref.~\cite{Motta:2021xwo} together with the relevant TOV. The radius constraints come from the works of \cite{Riley:2019yda}, labelled NICER1, \cite{Nicer2Miller_2019} which was labelled NICER 2, and the gravitational wave measurements \cite{ligoAbbott:2018exr} GW170817.}\label{bands}
	\end{figure}
\end{center}

 However, as shown by Motta {\em et al.}~\cite{Motta:2021xwo}, these characteristics are also shared by the quark-meson EoS discussed in this review, in particular that of the QMC model {\em once hyperons are included}.
In Fig.~\ref{bands} we can see the allowed EoSs from Ref.~\cite{Annala:2019puf} and the respective restrictions on the speed of sound. As discussed above, the QMC model yields a very similar reduction in the speed of sound to that given by quark matter.
The quark-meson based equations of state also respect the constraints of tidal deformability taken from the GW170817 measurement \cite{ligoAbbott:2018exr} and in every other respect they appear to be very similar to equations of state that suffer a soft transition to pure deconfined quark matter at high 
densities~\cite{Motta:2021xwo}. At  the moment, there appears to be no reliable way of determining if the EoS of neutron star matter does involve deconfinement at higher density. 

\section{Conclusion}
The composition of neutron stars is certainly still a major puzzle with no universally accepted answer. One of the most pressing issues is whether the baryonic matter suffers a phase transition to a deconfined state. Whether or not that happens at some high density, there can be no doubt that the role of changes in the internal structure of the baryons deserves serious attention. As discussed in section~\ref{finitenuclei}, there is ample evidence that the quark structure of the nucleon plays a role even in finite nuclei, perhaps providing an explanation of the EMC effect\cite{Thomas:1989vt}. The success of the QMC model \cite{Stone:2016qmi,Martinez:2018xep,Martinez:2020ctv} in describing heavy nuclei with so few model parameters is certainly an indication, albeit indirect, that such effects may be present. 

Neutron stars, with central densities many times the saturation density of normal nuclear matter should be expected to show even larger effects of the anticipated changes in the internal structure of baryons. Such studies are made particularly interesting by the appearance of hyperons in the cores of heavier stars. While the so-called ''hyperon puzzle'' has been resolved, other aspects, such as their role in neutrino cooling are just beginning to be studied seriously -- see for example Ref.~\cite{Anzuini:2021rjv}. Their participation in dense matter at the higher temperatures associated with supernovae and the proto-neutron stars formed after neutron star mergers is also a rich area for further study.

With new data from the NICER mission and from further gravitational wave measurements of neutron star mergers, we expect to soon be able to discriminate between the many models for the EoS of dense matter that we have discussed. 

\section*{Acknowledgements}
We are grateful to our many colleagues for their collaboration in projects related to this work, especially Pierre Guichon and Jirina Stone. This work was supported by the University of Adelaide and by the Australian Research Council through Discovery Project DP180100497 and the ARC Centre of Excellence for Dark Matter Particle Physics CE200100008.

\bibliographystyle{unsrtnat}
\bibliography{2021review.bib,jirina.bib}

\end{document}